\documentclass[twocolumn]{aastex61}

\received{2017 August 14}
\revised{2017 September 9}
\accepted{2017 September 13}
\submitjournal{ApJL}

\shorttitle{S-Cam Search of $z\sim6$ Quasars}
\shortauthors{Onoue et al.}

\begin{document}

\title{Minor Contribution of Quasars to Ionizing Photon Budget at $z\sim6$: Update on Quasar Luminosity Function at the Faint-end with Subaru/Suprime-Cam}

\author{Masafusa Onoue}
\email{masafusa.onoue@nao.ac.jp}
\affil{Department of Astronomical Science, 
Graduate University for Advanced Studies (SOKENDAI),
 2-21-1, Osawa, Mitaka, Tokyo 181-8588, Japan}
\affiliation{National Astronomical Observatory of Japan, 
2-21-1, Osawa, Mitaka, Tokyo 181-8588, Japan}

\author{Nobunari Kashikawa}
\affil{Department of Astronomical Science, 
Graduate University for Advanced Studies (SOKENDAI),
 2-21-1, Osawa, Mitaka, Tokyo 181-8588, Japan}
\affiliation{National Astronomical Observatory of Japan, 
2-21-1, Osawa, Mitaka, Tokyo 181-8588, Japan}

\author{Chris J. Willott}
\affil{Herzberg Institute of Astrophysics, 
National Research Council, 5071 West Saanich Road,
Victoria, BC V9E 2E7, Canada}

\author{Pascale Hibon}
\affil{European Southern Observatory, Alonso de Cordova 3107, 
Casilla 19001, Santiago, Chile}

\author{Myungshin Im}
\affil{Center for the Exploration of the Origin of the Universe (CEOU),
Astronomy Program,
Department of Physics and Astronomy, 
Seoul National University, 1-Gwanak-rho,
Gwanak-gu, Seoul 141-742, Korea}

\author{Hisanori Furusawa}
\affiliation{National Astronomical Observatory of Japan, 
2-21-1, Osawa, Mitaka, Tokyo 181-8588, Japan}

\author{Yuichi Harikane}
\affiliation{Institute for Cosmic Ray Research, The University of Tokyo,
5-1-5, Kashiwanoha, Kashiwa, Chiba 277-8582, Japan}
\affiliation{Department of Physics, Graduate School of Science,
The University of Tokyo, Bunkyo, Tokyo 113-0033, Japan}

\author{Masatoshi Imanishi}
\affil{Department of Astronomical Science, 
Graduate University for Advanced Studies (SOKENDAI),
 2-21-1, Osawa, Mitaka, Tokyo 181-8588, Japan}
\affiliation{National Astronomical Observatory of Japan, 
2-21-1, Osawa, Mitaka, Tokyo 181-8588, Japan}

\author{Shogo Ishikawa}
\affiliation{National Astronomical Observatory of Japan, 
2-21-1, Osawa, Mitaka, Tokyo 181-8588, Japan}

\author{Satoshi Kikuta}
\affil{Department of Astronomical Science, 
Graduate University for Advanced Studies (SOKENDAI),
 2-21-1, Osawa, Mitaka, Tokyo 181-8588, Japan}
\affiliation{National Astronomical Observatory of Japan, 
2-21-1, Osawa, Mitaka, Tokyo 181-8588, Japan}

\author{Yoshiki Matsuoka}
\affiliation{Research Center for Space and Cosmic Evolution,
Ehime University, Matsuyama, Ehime 790-8577, Japan}


\author{Tohru Nagao}
\affiliation{Research Center for Space and Cosmic Evolution,
Ehime University, Matsuyama, Ehime 790-8577, Japan}

\author{Yuu Niino}
\affiliation{National Astronomical Observatory of Japan, 
2-21-1, Osawa, Mitaka, Tokyo 181-8588, Japan}

\author{Yoshiaki Ono}
\affiliation{Institute for Cosmic Ray Research, The University of Tokyo,
5-1-5, Kashiwanoha, Kashiwa, Chiba 277-8582, Japan}

\author{Masami Ouchi}
\affiliation{Institute for Cosmic Ray Research, The University of Tokyo,
5-1-5, Kashiwanoha, Kashiwa, Chiba 277-8582, Japan}

\author{Masayuki Tanaka}
\affiliation{National Astronomical Observatory of Japan, 
2-21-1, Osawa, Mitaka, Tokyo 181-8588, Japan}

\author{Ji-Jia Tang}
\affil{Institute of Astronomy and Astrophysics, Academia Sinica, 
Taipei, 10617, Taiwan}

\author{Jun Toshikawa}
\affiliation{Institute for Cosmic Ray Research, The University of Tokyo,
5-1-5, Kashiwanoha, Kashiwa, Chiba 277-8582, Japan}

\author{Hisakazu Uchiyama}
\affil{Department of Astronomical Science, 
Graduate University for Advanced Studies (SOKENDAI),
 2-21-1, Osawa, Mitaka, Tokyo 181-8588, Japan}
\affiliation{National Astronomical Observatory of Japan, 
2-21-1, Osawa, Mitaka, Tokyo 181-8588, Japan}



\begin{abstract}

We constrain the quasar contribution to cosmic reionization based on our deep optical survey of $z\sim6$ quasars down to $z_R=24.15$ using Subaru/Suprime-Cam in three UKIDSS-DXS fields covering $6.5$ deg$^2$.
In \citet{Kashikawa15}, we select 17 quasar candidates and report our initial discovery of two low-luminosity quasars ($M_{1450}\sim-23$) from seven targets,
one of which might be a Ly$\alpha$ emitting galaxy.
From an additional optical spectroscopy, 
none of the four candidates out of the remaining ten turn out to be genuine quasars.
Moreover, the deeper optical photometry provided by the Hyper Suprime-Cam Subaru Strategic Program (HSC-SSP) shows that, unlike the two already-known quasars, the $i-z$ and $z-y$ colors of the last six candidates are consistent with M- or L-type brown dwarfs.
Therefore, the quasar luminosity function (QLF) in the previous paper is confirmed.
Compiling QLF measurements from the literature over a wide magnitude range, including an extremely faint 
AGN candidate from \citet{Parsa17}, to fit them with a double power-law, we find that the best-fit faint-end slope is $\alpha=-2.04^{+0.33}_{-0.18}$ ($-1.98^{+0.48}_{-0.21}$) and characteristic magnitude is $M_{1450}^{*}=-25.8^{+1.1}_{-1.9}$ ($-25.7^{+1.0}_{-1.8}$) in the case of two (one) quasar detection.
Our result suggests that, if the QLF is integrated down to $M_{1450}=-18$, quasars produce $\sim1-12$\% of the ionizing photons required to ionize the whole universe at $z\sim6$ with $2\sigma$ confidence level, assuming that the escape fraction is $f_\mathrm{esc}=1$ and the IGM clumpy factor is $C=3$. 
Even when the systematic uncertainties are taken into account,
our result supports the scenario that quasars are the minor contributors of reionization.
\end{abstract}




\section{Introduction} \label{sec:intro}
High-redshift ($z\gtrsim6$) quasars are unique probes of the early universe in a way complementary to other populations such as galaxies and gamma-ray bursts.
In particular, they can be used to estimate the hydrogen neutral fraction of the intergalactic medium (IGM), which directly probes the cosmic reionization history \citep{Fan06A}.
Furthermore, the existence of the most massive ($M_\mathrm{BH}>10^{9}M_\odot$) supermassive black holes at this epoch poses a challenge to the seed black hole formation and early growth scenario \cite[e.g.,][]{Mortlock11, Wu15}.

Searches of $z>5.7$ quasars have been performed in large optical and near-infrared surveys, namely SDSS \citep{Fan06,Jiang16}, CFHQS \citep{Willott10b}, UKIDSS \citep{Mortlock11}, VIKING \citep{Venemans13}, DECaLS \citep{Wang17}, DES \citep{Reed17}, PS1 \citep{Banados16}, and HSC \citep{Matsuoka16, Matsuoka17}, 
which have identified more than two hundred quasars to date.
Thanks to the large sample size, the bright end of the quasar luminosity function (QLF) is well constrained \citep[e.g.,][]{Jiang16}.
However, the photon budget of quasars, i.e., how much ionizing photons quasars emit to reionize the universe is still an outstanding issue as the QLF faint-end, which is crucial in the photon budget estimate, is poorly constrained.
It has been based on three (or two) quasars at the faintest range ($M_{1450}\sim-22$) which are spectroscopically identified in \citet{Willott10b} and \citet[][hereafter K15]{Kashikawa15}.
Whereas these optical studies suggest insufficient contribution of quasars, \citet{Giallongo15} propose a scenario that faint AGN are the major contributors in the ionizing background radiation at $4<z<6$,
based on their 22 extremely faint ($M_{1450}\sim-20$) AGN candidates in the GOODS-South field, five of which are at $z\sim5.75$.
Their X-ray fluxes are detected in deep X-ray 4Ms {\it Chandra} images.
Following this paper and the discovery of a long and dark Lyman alpha trough in a $z\simeq6$ quasar spectrum \citep{Becker15, Chardin17}, such AGN-driven reionization scenario has recently been vigorously discussed in, for example, \citet{Madau15}.
However, \citet{Parsa17} show that, 
through their individual examination of the X-ray images and photometric redshift considering a wide range of dust reddening in their galaxy and AGN template SEDs, only seven of the $z>4$ AGN candidates are robust, among which one is at $z>5$.
Several other deep X-ray studies also report discrepancy with the \citet{Giallongo15} results \citep{Vito16, Ricci17}.
The photon budget issue has also been discussed with the UV luminosity function of Lyman break galaxies \citep[e.g.,][]{Finkelstein15, Bouwens16}, which is well constrained down to $M_\mathrm{UV}\sim -13$.
However, unclear understanding of the escape fraction of UV photons and the magnification uncertainties in the gravitational lensing prevent one from making a convincing conclusion on which population, galaxies or AGN is the dominant contributor of the reionization.
This argument is particularly essential for estimating the typical size of the ionizing bubble, which essentially depends on the ionizing sources.

In this {\it letter}, we update our deep survey of $z\sim6$ quasars with Subaru/Suprime-Cam \citep{Miyazaki02}, the initial results of which are reported in K15.
We adopt a standard $\Lambda$CDM cosmology with $H_0=70$ km s$^{-1}$ Mpc$^{-1}$, $\Omega_m=0.3$, $\Omega_\Lambda=0.7$, and $\Omega_bh^2=0.022$.
Magnitudes are given in the AB system. 

\section{Subaru/Suprime-Cam Observation and Previous Results} \label{sec:previous}
We observed $6.5$ deg$^2$ in total consisting of three UKIDSS-DXS fields (Lockman Hole, ELAIS-N1, VIMOS~4) on June 22--24, 2009 (UT) exploiting two broad bands of the Suprime-Cam: $z_B$ ($\lambda_\mathrm{eff}=8842$\AA) and $z_R$ ($\lambda_\mathrm{eff}=9841$\AA) with $3\sigma$ depth\footnote{$2\arcsec\phi$-aperture limiting magnitude } of $25.55$ and $24.15$, respectively.
CFHTLS $i'$-band ($\lambda_\mathrm{eff}=7571$\AA) images are provided in VIMOS~4 field with $i'_{\mathrm{lim},3\sigma}\sim 25.7$\footnote{
See Section~\ref{sec:phot_difference}.
}.
Suprime-Cam $i'$-band ($\lambda_\mathrm{eff}=7641$\AA) images are available in Lockman/ELAIS fields with $i'_{\mathrm{lim},3\sigma}\sim 26.42$.
UKIDSS $J$- and $K$-band images are also available over the three fields with $J_{\mathrm{lim},3\sigma}=23.84$ and $K_{\mathrm{lim},3\sigma}=23.17$. 
In K15, we select $17$ candidates at $23.65\leq z_R\leq24.11$ from stellar objects applying following color selection: i) $i'-z_B>1.7$, ii) $z_B - z_R<1.0$, iii) $i'-z_B > 2(z_B - z_R)+0.9$,
which can effectively distinguish quasars from contaminants such as Galactic brown dwarfs.
We ignore the difference of the two $i'$-band filters causing $\Delta i'<0.01$ mag difference for a $z\sim6$ quasar.
The observed magnitudes and coordinates of the candidates are listed in Table~3 of K15 and Table \ref{tab:phot} of this paper with new photometry.
From our initial follow-up spectroscopy with Subaru/FOCAS \citep{Kashikawa02}, we discovered two $z\sim6$ quasars out of seven targets, giving an initial constraint on the QLF faint-end.
That paper suggests that quasars are responsible for $5-15$\% of the ionizing photons at $z\sim6$, but we could only give a lower-limit due to the remaining ten candidates yet to be identified.
Note that ELAIS1091000446 ($z_R=24.2$) at $z=6.04$, one of the discovered quasars has an unusual Ly$\alpha$ profile with its half-line width of $427$ km s$^{-1}$, which falls between those of typical quasars ($>1000$ km s$^{-1}$) and \replaced{Ly$\alpha$ emitters ($\sim100-200$ km s$^{-1}$)}{Lyman break galaxies ($\sim100$ km s$^{-1}$)}.

\section{Follow-up Spectroscopy} \label{sec:spec_obs}
We performed optical spectroscopy for four of the remaining candidates (VIMOS2873003200, ELAIS891006630, ELAIS914002066 and ELAIS914003931) with Subaru/FOCAS (S15B-204S \& S16A-200S\deleted{, PI: M.Onoue}).
We selected the targets with relatively red $i'-z_B$ colors among the ten.
Our program was executed on September 8, 2015 \deleted{for the first two targets} and May 14, 2016 (UT). \deleted{for the last two}
We used 300R grism with an order-cut filter O58 to cover $5800-10450$ \AA\  with a spectral resolution of $R\sim 500$ ($0\arcsec.8$-slit width).
Each target was observed for about two hours, divided into six $1020$ sec (September run) and $1077$ sec (May run) exposures.
Within the six exposures, three-point dithering ($-1\arcsec.0$, $0\arcsec.0$, $+1\arcsec.0$) was applied for good background subtraction.
Two-pixel CCD on-chip binning in the spatial direction ($0\arcsec.104$ per pix) was applied.
The sky conditions were clear with $0\arcsec.4-0\arcsec.7$ seeing size.
The data was reduced with a standard procedure using an IRAF-based FOCAS pipeline.
The individual 2D spectra were median stacked and reduced to final 1D spectra.

As a result, we find that none of the four targets are quasars.
The continuum flux of VIMOS2873003200 ($z_R=23.78$) is detected all over the spectral coverage with a doublet emission line at $\lambda_\mathrm{obs}\sim8770$ \AA, which is likely a [O{\sc ii}] doublet $\lambda\lambda 3726\  3729$ at $z=1.354$.
Note that another [O{\sc ii}] emitter (Lockman14004800) was found in K15.
The other three ELAIS targets are not detected.
As we describe in Section~\ref{sec:phot_difference}, these are also undetected in deeper photometric observation with Subaru/HSC.

\section{The HSC Colors} \label{sec:phot_obs}
\subsection{Last Six Candidates} \label{sec:phot_last}
This Suprime-Cam survey of $z\sim6$ quasars was originally positioned as a pre-study for a much larger survey, the Hyper Suprime-Cam Subaru Strategic Program (HSC-SSP), of which
the detailed design and the survey strategy are given in \citet{DRpaper, SSPpaper}.
Fortunately, ELAIS-N1 and VIMOS~4 fields are covered in the HSC-Deep and Wide layers, respectively.
We leverage the latest internal photometric catalog of the HSC-SSP (DR S16A) to inspect the last six candidates.
The average $5\sigma$ depths of the PSF magnitude in Wide (Deep) are $g\sim26.8$ ($\sim26.8$), $r\sim26.4$ ($\sim26.6$), $i\sim26.4$ ($\sim26.5$), $z\sim25.5$ ($\sim25.6$), and $y\sim24.7$ ($\sim24.8$).
The filter information of the HSC is described in Kawanomoto et al. (in prep.).

\begin{figure}[tb]
\plotone{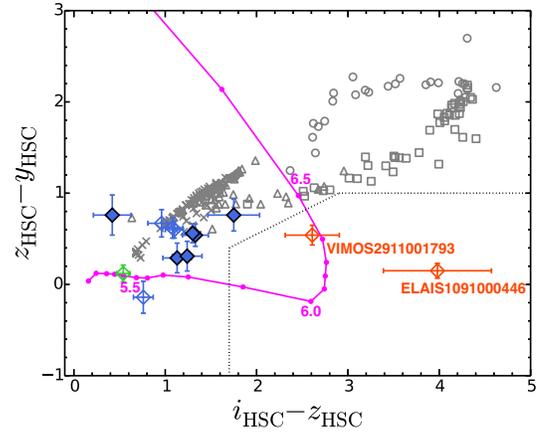}
\caption{HSC $i-z$ and $z-y$ color diagram of the 13 quasar candidates detected in the HSC-SSP.
The red, green, and blue diamonds show the quasars (VIMOS2911001793, ELAIS1091000446), [O{\sc ii}] emitter, and other sources, respectively.
Spectroscopically-identified candidates are shown in open symbols and the other six are shown in filled symbols.
The model color track of $5<z<7$ quasars are shown in a solid line with steps of $\Delta z=0.1$.
The model colors of M (cross), L (triangle), T (square), and Y-type (circle) brown dwarfs from the BTsettl model \citep{BTsettl} are shown in grey.
The color selection window applied in K15 is shown in a dotted line.
\label{fig:2color}}
\end{figure}
Figure~\ref{fig:2color} shows the $i-z$ and $z-y$ color diagram of $13$ candidates detected in the HSC-SSP.
The solid line shows the quasar color track at $5<z<7$ and the grey symbols show the colors of brown dwarfs, which are derived from the SED templates of the BTsettl model \citep{BTsettl}\footnote{Solar metallicity models are used.}.
As a result, we find that, using the same color selection used in K15, only the two already-identified quasars have strong color excess in $i-z$.
One of the six unclassified candidates, VIMOS2773005145 shows moderately-red colors in $i-z$ and  $z-y$, which are indicative of \deleted{a smoothly increasing continuum of} a L-type brown dwarf \deleted{at $0.7<\lambda_\mathrm{obs}<1.0$ $\mu$m}.
The $i-z$ colors of the other five are not as red as expected for $z\sim6$ quasars ($0.4\leq i-z\leq1.3$) and rather consistent with M-type brown dwarfs, which is also the case for other non-quasars spectroscopically followed-up in K15\footnote{VIMOS2871007103 is likely a low-$z$ interloper as it is detected in $g$ and $r$. See Table~\ref{tab:phot}.}. \deleted{(blue open diamonds)}
\deleted{Moreover, the three ELAIS candidates undetected in the FOCAS spectroscopy (Sec.~\ref{sec:spec_obs}) are also undetected in the HSC-SSP photometric catalog, although the imaging is deeper than the Suprime-Cam observation.
Therefore, it is likely that they are either moving or transient objects only bright at the detected positions during our Suprime-Cam observation.
The [O{\sc ii}] emitter found in this paper, VIMOS2873003200 is also detected in bluer $g$- and $r$-bands with $>10\sigma$ significance.}
Conclusively, the deeper HSC photometry shows that there are no additional quasars in our remaining candidates.
The HSC photometry is summarized in Table~\ref{tab:phot}.

\subsection{The Difference of S-Cam \& HSC Colors} \label{sec:phot_difference}
We suspect that the different colors of the candidates between the S-Cam+CFHT and HSC filter sets are caused by different imaging depths.
The HSC-Wide is deeper by $0.5$ and $1.3$ mag in $i$ (S-Cam and CFHT $i'$, respectively), $0.5$ mag in $z$ ($z_B$) and $1.1$ mag in $y$ ($z_R$).
While the candidates are detected in $z_B$ and $z_R$ with more than $7\sigma$ significance, they are faint in $i'$ with $\lesssim1\sigma$ detections.
Therefore, it can be said that the deeper imaging of the HSC helps derive robust optical colors of the candidates.
Moreover, we find that our previous estimate of the limiting magnitude of $i'$-band was slightly optimistic in the VIMOS field. 
We re-evaluated the $i'$-band depth based on the distribution of background levels within $2\arcsec\phi$-apertures at randomly selected positions on the images.
The depth was found to be systematically shallower by $\sim1$ mag than the one in K15, with a large field-to-field variation of $0.6$ mag partly due to heavy galactic cirrus \citep[$A_V\sim0.16$ mag,][]{Schlafly11}.
The total effective volume is evaluated to be at most $\sim25$\% smaller than in K15, given their selection completeness.

\added{The three ELAIS candidates undetected in the FOCAS spectroscopy (Secion~\ref{sec:spec_obs}) are also undetected in the HSC-SSP images: $\gtrsim 3$ mag fainter than in $z_B$ and $z_R$.
Considering the rest-frame UV variability of quasars ($\lesssim1$ mag, \citealp{Welsh11}), they are likely to be either moving or transient objects in the Suprime-Cam images taken on continuous days, rather than variable quasars.
The [O{\sc ii}] emitter found in this paper, VIMOS2873003200 is also detected in bluer $g$- and $r$-bands with $>10\sigma$ significance.
Note that there are other quasar candidates in ELAIS and VIMOS at the similar magnitude range of the K15 candidates, if we only use the deeper HSC colors.
However, it is a natural consequence considering the low completeness in our Supreme-Cam selection ($\lesssim50$\%).
We will discuss them in the series of the HSC quasar papers \citep{Matsuoka16, Matsuoka17, Akiyama17}.}


\begin{deluxetable*}{lCCCCCl}
\tablecaption{\deleted{S-Cam \&} HSC photometry of 17 candidates \label{tab:phot}}
\tablecolumns{7}
\tablenum{1}
\tablewidth{0pt}
\tablehead{
\colhead{Object}  &\colhead{$g$} &\colhead{$r$} &\colhead{$i$} &\colhead{$z$} &\colhead{$y$} & \colhead{Notes} 
}
\startdata
\multicolumn{7}{c}{Candidates followed-up in Kashikawa+15}\\
\hline
VIMOS2911001793& >28.5 & >28.1 & 27.49\pm0.59& 23.51\pm0.04& 23.36\pm0.07& $z=6.156$ QSO\\
ELAIS1091000446 & 28.56\pm0.67 & 27.41\pm0.54 & 27.03\pm0.29& 24.42\pm0.06& 23.88\pm0.09& $z=6.041$ QSO\\
Lockman14004800 & \cdots &\cdots&\cdots&\cdots&\cdots& [O{\sc ii}] emitter\\
VIMOS2832005555& >28.5&27.96\pm1.08&25.09\pm0.07& 23.99\pm0.07& 23.39\pm0.06& ND\\
VIMOS2871008551& >28.5& >28.1&25.63\pm0.09& 24.67\pm0.12& 24.00\pm0.09& ND\\
VIMOS2871007103& 25.92\pm0.07&25.64\pm0.11&25.09\pm0.06& 24.33\pm0.09& 24.47\pm0.15& ND\\
VIMOS3031005637& >28.5&26.53\pm0.28&25.40\pm0.08& 24.32\pm0.08& 23.70\pm0.08& ND\\
\hline
\multicolumn{7}{c}{Remaining candidates}\\
\hline
VIMOS2873003200 & 24.74\pm0.03&24.56\pm0.04&24.40\pm0.04& 23.86\pm0.06& 23.74\pm0.07& [O{\sc ii}] emitter\\
ELAIS891006630 & >28.5&>28.3&>28.2&>27.3&>26.5& ND\\
ELAIS914002066 & >28.5&>28.3&>28.2&>27.3&>26.5& ND\\
ELAIS914003931 & >28.5&>28.3&>28.2&>27.3&>26.5& ND\\
VIMOS2752003989 & >28.5&27.79\pm0.79&25.54\pm0.10& 24.21\pm0.10& 23.67\pm0.07& likely BD\\
VIMOS2773005145 & >28.5&>28.1&26.35\pm0.24& 24.60\pm0.15& 23.84\pm 0.09& likely BD\\
VIMOS2833009245 & >28.5&>28.1&25.64\pm0.11& 24.51\pm0.11& 24.22\pm0.12& likely BD\\
VIMOS2853001577 & >28.5&>28.1&25.26\pm0.08& 24.84\pm0.19& 24.08\pm0.11& likely BD\\
VIMOS2733006446 & >28.5&>28.1&25.76\pm0.11& 24.52\pm0.12& 24.21\pm0.11& likely BD\\
VIMOS2993006408 & >28.5&>28.1&25.52\pm0.07& 24.22\pm0.05& 23.66\pm0.09& likely BD\\
\enddata
\tablecomments{$1\sigma$ limiting magnitude is shown for undetected bands. 
Lockman field is not covered in the HSC-SSP.
In the last column, candidates undetected in the spectroscopic follow-up are shown as ``ND", while candidates not targeted in the spectroscopy and showing stellar-like HSC colors are shown as ``likely BD".
\added{The S-Cam photometry can be found in K15.}
}
\end{deluxetable*}

\section{Discussion}\label{sec:discussion}
%
 
\subsection{$z\sim6$ QLF} \label{sec:QLF}
\begin{figure*}[tb!]
\plotone{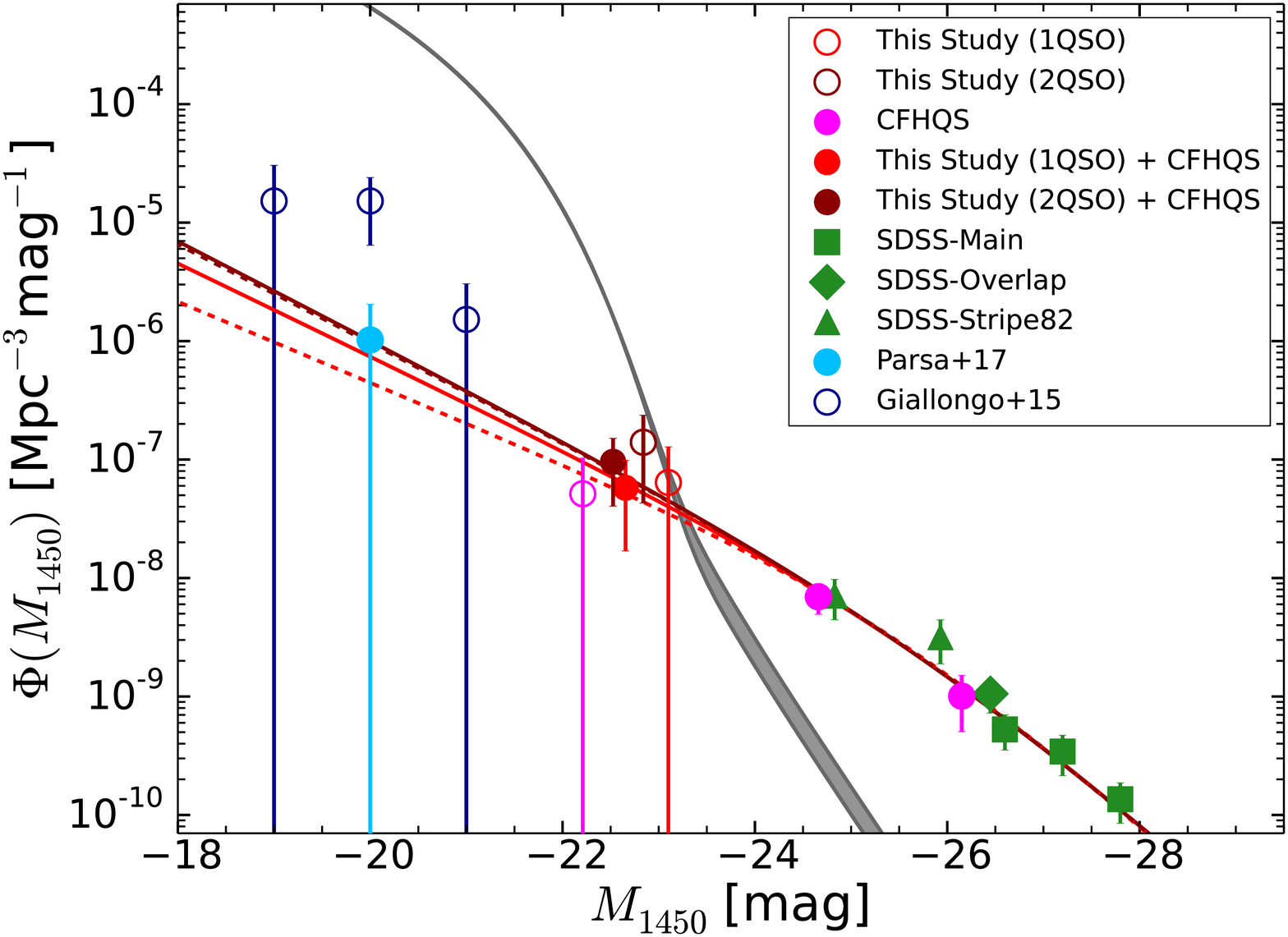}
\caption{$z\sim6$ QLF. The binned QLF from K15 and this study is shown in brown (case 1) and red (case 2) open symbols, respectively, with the re-binned QLFs with the faintest bin of the CFHQS \citep[][magenta]{Willott10b} shown in filled symbols. 
The SDSS data points (Main, Overlap, Stripe82) from \citet{Jiang16} are shown in green (squares, diamonds, and triangles).
AGN luminosity functions from \citet[][cyan]{Parsa17} and \citet[][blue]{Giallongo15} are scaled to $z=6.0$ using $k=-0.47$.
The best-fit double power-laws (using all the filled symbols) are shown in solid lines, while those fitted without the \citet{Parsa17} bin are shown in dashed lines.
The shaded grey line shows the UV luminosity function of $i$-dropout galaxies from \citet[][lensed Schechter function]{Ono17}, which intersects with the QLFs at $M_{1450}\simeq -23$. \label{fig:qlf}}
\end{figure*}

As we find no additional quasars from the remaining candidates, our previous constraint on the binned QLF in K15 is confirmed.
In Figure~\ref{fig:qlf}, the binned QLF from our Suprime-Cam survey (K15 and this study) is shown in brown (two quasars) and red (one quasar) open circles with Poisson error bars.
We do not correct for the effect of the shallow imaging in the VIMOS field, because we find that it only increases the ionizing photon emissivity of quasars by $\lesssim6$\%, which is negligible within our QLF constraints.
We also show the binned QLF of the CFHQS \citep[][magenta]{Willott10b} and those of the SDSS \citep[][green]{Jiang16} in squares (SDSS-Main), triangles (Overlap region), and diamonds (Stripe 82).
The AGN luminosity functions derived in \citet[][blue]{Giallongo15} and \citet[][cyan]{Parsa17} are also shown at $-21\leq M_{1450}\leq-19$, while it should be noted that their luminosity functions are based on AGN ``candidates" with large uncertainties on the photometric redshift (see Parsa paper for detailed discussions) and host galaxy contribution may not be negligible in the rest-frame UV luminosity for such ultra-faint sources \citep{Ricci17}.
To calculate the shape of the $z\sim6$ QLF, $\Phi(M_{1450}, z)$, we fit the binned QLF with a double power-law:
\begin{equation}
\Phi(M_{1450},z)=\frac{10^{k(z-6)}\Phi^*(z)}{10^{0.4(\alpha+1)(M_{1450}-M_{1450}^*)}+10^{0.4(\beta+1)(M_{1450}-M_{1450}^*)}},
\end{equation}
where $\alpha$ and $\beta$ are the faint- and bright-end slopes, respectively, $M_{1450}^*$ is the characteristic magnitude, and $k$ is the parameter showing the redshift evolution.
The scale factor $\Phi^*(z)$ is a function of redshift.
We fix the bright-end slope as $\beta=-2.8$ following the SDSS result \citep{Jiang16} since our data point is at the faint side.
The LFs of \citet{Giallongo15} and \citet{Parsa17} at $z=5.75$ are scaled to $z=6.0$ using $k=-0.47$, which is derived in \citet{Fan01}.
Since the faintest bins of \citet{Willott10b} and ours are based on discovery of only one ($M_{1450}=-22.21$) and two quasars ($M_{1450}=-23.10, -22.58$), we combine the two individually determined bins into one as also applied in K15.
These re-binned QLFs are shown in filled circles in Figure~\ref{fig:qlf} and used in the following analysis.
We assume two cases in which two quasars are detected  (case 1), and one quasar is detected (case 2) in our survey due to the unclear classification of ELAIS1091000446.
In order to estimate the best-fit combination of ($\alpha$, $M_{1450}^*$), we fit the double power-law using $\chi^2$ minimization to the binned QLFs shown in filled symbols in Figure~\ref{fig:qlf}, namely the SDSS, CFHQS, our study (re-binned with CFHQS), and the X-ray selected AGN candidate from \citet{Parsa17}.
Note that the scale-factor, $\Phi^*(z=6)$ is determined so that the minimum $\chi^2$ is achieved at each ($\alpha$, $M_{1450}^*$).

\begin{figure}[tb!]
\plotone{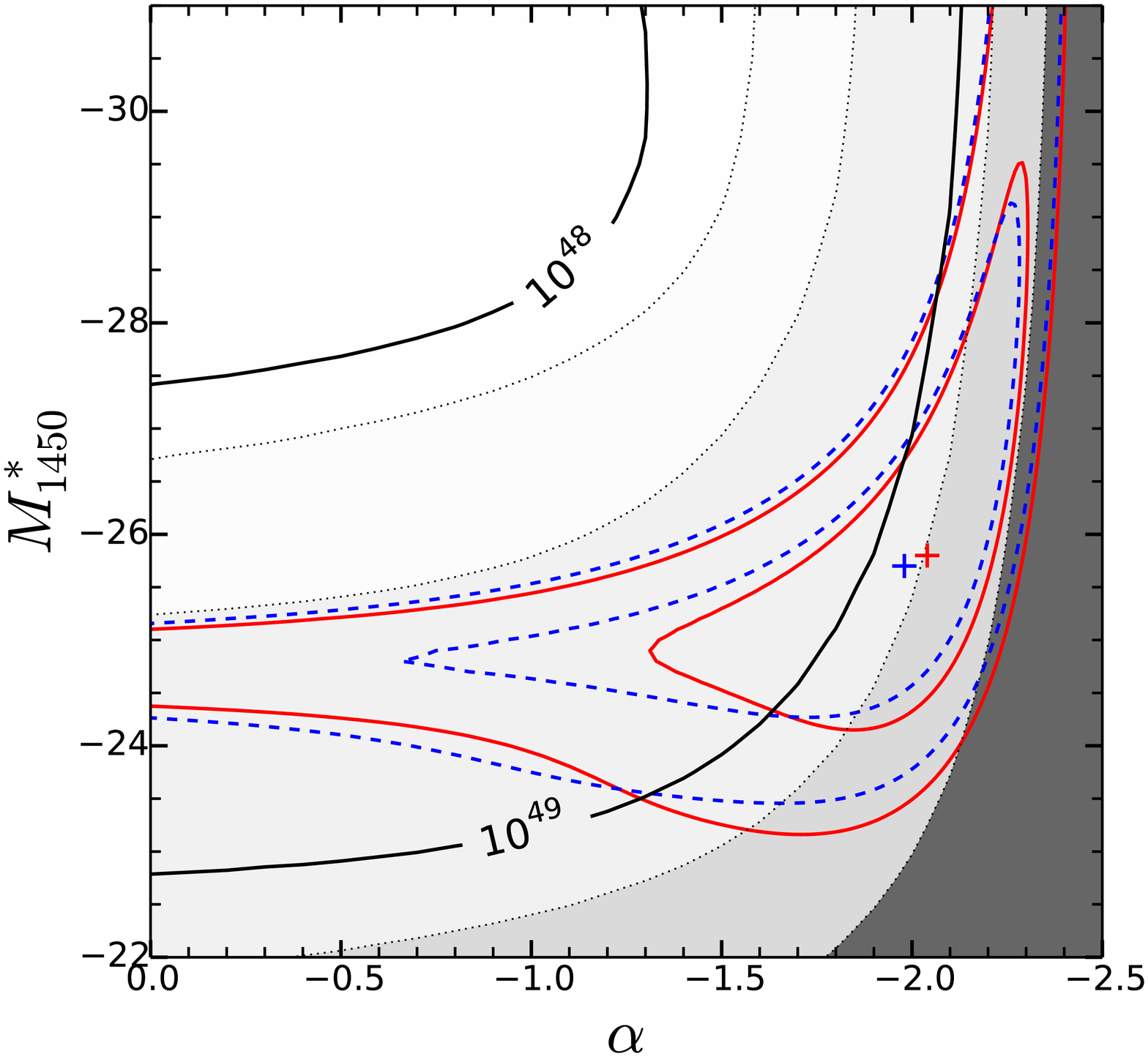}
\caption{Constraints on the QLF faint-end slope $\alpha$ and characteristic absolute magnitude $M_{1450}^*$.
The $1\sigma$ and $2\sigma$ confidence ranges 
are shown in red and blue contours, respectively, with the best-fit indicated in plus signs.
The ionizing emissivity of quasars, $\dot{N}_\mathrm{ion}$ [s$^{-1}$ Mpc$^{-3}$] for each ($\alpha$, $M_{1450}^*$) is indicated in solid lines assuming case 1
and the corresponding photon budget is shown in shades ($10, 5, 1, 0.5$\% from right to left).\label{fig:pb}}
\end{figure}

As a result, we derive ($\alpha$, $M_{1450}^*$) $=$ ($-1.63^{+1.21}_{-1.09}$, $-25.8^{+1.1}_{-1.9}$) with the scale factor $\Phi^*(z=6)=4.06\times10^{-9}$ Mpc$^{-3}$ mag$^{-1}$ for the case 1 and ($\alpha$, $M_{1450}^*$) $=$ ($-1.98^{+0.48}_{-0.21}$, $-25.7^{+1.0}_{-1.8}$) with the scale factor $\Phi^*(z=6)=4.53\times10^{-9}$  Mpc$^{-3}$ mag$^{-1}$ for the case 2.
These best-fit QLFs are shown in solid lines in Figure~\ref{fig:qlf}.
In both cases, the best-fit parameters are consistent with the QLF in \citet{Jiang16} within $1\sigma$ level, while they also use the K15 and \citet{Willott10b} results (assuming one quasar in K15). 
To estimate the uncertainty of the faint-end slope and the characteristic magnitude, we fit the QLF with fixed ($\alpha$, $M_{1450}^*$) over $-2.50\leq\alpha\leq 0.00$ and $-31.0\leq M_{1450}^*\leq-22.0$ with steps of $\Delta\alpha=0.01$ and $\Delta M_{1450}=0.1$.
In Figure~\ref{fig:pb}, the two-dimensional $1\sigma$ and $2\sigma$ confidential ranges are shown in red (case 1) and blue (case 2) contours with the best-fit values.
We also fit the QLFs excluding the \citet{Parsa17} bin, and find that the number densitiy at the faint-end slightly decreases, which are indicated in dashed lines in Figure~\ref{fig:qlf}.
Table~\ref{tab:pb_summary} summarizes our best-fit QLF parameters.

We repeat the QLF fitting by replacing the \citet{Parsa17} bin with those of \citet{Giallongo15}, assuming two quasars from our survey. 
The derived QLF is not strongly different from the above results, and the faint-end slope is required to be much steeper ($\alpha=-2.15$) than their measurement ($\alpha=-1.66$).
Note that the best-fit power-law falls at smaller densities than \citet{Giallongo15} in all bins, while the goodness-of-fit is reasonable ($\chi^2_\nu=0.9$) partly due to the large uncertainties in their constraints.
In addition, the total number density of quasars when the QLF is integrated down to $M_{1450}=-18$ is only $25\%$ of theirs.

When our best-fit QLFs are compared to lower redshifts, there is a trend that, albeit with a large uncertainty, the faint-end slope gets steeper from $z\sim4$ \citep[$\alpha=-1.30\pm0.05$,][]{Akiyama17}, which is indicative of the supermassive black holes in their actively growing phase at $z>6$.
\deleted{The scale factor $\Phi(z)$ declines from $z\sim4$ by about two orders, while the characteristic magnitude $M^*_{1450}$ does not show strong evolution (see Figure~20 of Akiyama17).}

\subsection{Photon Budget of Quasars during Reionization} \label{sec:pb}
The ionizing photon emissivity of $z\sim6$ quasars is calculated from the QLFs derived in the previous section, following the framework given in \citet{BH07}. 
In this paper, we assume a broken power-law for the rest-frame SED of a quasar: $f_\nu\propto\nu^{-0.6}$ ($\lambda\leq912$\AA) and $f_\nu\propto\nu^{-1.7}$ ($\lambda_\mathrm{rest}>912$\AA), which is given in \citet{Lusso15}.
We integrate the QLF from $M_{1450}=-30$ to $-18$ mag using the best-fit double power-law at each frequency to derive the monochromatic emissivity.
\deleted{Then, it is converted to the hydrogen ionization rate assuming that all ionizing photons ($\lambda\leq912$\AA) emitted from quasars escape to the IGM, from which
the ionizing photon emissivity per unit comoving volume, $\dot{N}_\mathrm{ion}$ is derived.}
Based on our best-fit QLFs, we find $\dot{N}_\mathrm{ion}=1.63\times10^{49}$ s$^{-1}$ Mpc$^{-3}$ (case 1) and $\dot{N}_\mathrm{ion}=1.34\times10^{49}$ s$^{-1}$ Mpc$^{-3}$ (case 2).
The required emission rate to balance with the hydrogen recombination at $z=6$ is $\dot{N}_\mathrm{ion}=10^{50.48}\ (3/C)$ s$^{-1}$ Mpc$^{-3}$ \citep{Madau99}, where $C$ is the IGM clumping factor.
Therefore, the contribution of $z\sim6$ quasars to the ionizing photons is $5.4$\%  in case 1 and $4.4$\%  in case 2, if we assume the fiducial value of $C=3$ \citep[e.g.,][]{Shull12}.
In Figure~\ref{fig:pb}, the ionizing photon emissivity at each $\alpha$ -- $M_{1450}^*$ plane is indicated with solid lines assuming case 1.
The shades show the corresponding photon budget,
which indicates that the $2\sigma$ confidence range falls at $\sim1-12$\% contribution in either case.
The error estimate is based on the two-dimensional confidence range of $\alpha$ and $M_{1450}^*$ (Fig.~\ref{fig:pb}).
Meanwhile, 
if the QLF is integrated further down to $M_{1450}=-13$, where the recent studies of $z>6$ UV luminosity function \added{of galaxies} exploiting gravitational lensing have reached \citep{Livermore17, Bouwens16},
the photon budget is still small ($9.8$\% and $7.0$\% for case 1 and 2, respectively).
Taking systematic uncertainties due to the choice of the clumping factor ($C\sim2-5$, \citealt{Jiang16}), the escape fraction, and the minimum magnitude of the QLF into account,
we place a stringent upper limit of $\lesssim20$\%; thus, our result supports the scenario that quasars are likely the minor contributors of the ionizing background at $z\sim6$.

\begin{deluxetable*}{lCCCCC}[htb]
\tablecaption{Best-fit QLF Parameters \label{tab:pb_summary}}
\tablecolumns{6}
\tablenum{2}
\tablewidth{0pt}
\tablehead{
\colhead{Case} &
\colhead{$\alpha$} &
\colhead{$M_{1450}^{*}$} &
\colhead{$\Phi^*(z=6)$} & 
\colhead{$\dot{N}_\mathrm{ion}^a$} & 
\colhead{$\chi_\nu^{2,b}$} \\
\colhead{} & \colhead{} &
\colhead{(mag)} & \colhead{ ($\times10^{-9}$ Mpc$^{-3}$ mag$^{-1}$)} &\colhead{($\times10^{49}$ s$^{-1}$ Mpc$^{-3}$)} & \colhead{}
}
\startdata
1 & -2.04^{+0.33}_{-0.18} & -25.8^{+1.1}_{-1.9} &4.06& 1.63  & 0.5\\
1' & -2.03^{+0.54}_{-0.32} & -25.7^{+1.1}_{-3.6} &4.25 & 1.48 & 0.6\\
2 & -1.98^{+0.48}_{-0.21} & -25.7^{+1.0}_{-1.8} &4.53 & 1.34 & 0.5\\
2' & -1.85^{+0.86}_{-0.40} & -25.4^{+0.9}_{-2.2} &6.53 & 1.00 & 0.6\\
\enddata
\tablecomments{
$^a$ Ionizing emissivity with corresponding photon budget in parentheses in which we assume that the escape fraction is $f_\mathrm{esc}=1$ and the IGM clumping factor is $C=3$.
The cases where the X-ray based bin of \citet{Parsa17} is excluded for case 1 and 2 are shown in case 1' and 2', respectively. 
$^b$ Reduced chisquare of the QLF fitting.}
\end{deluxetable*}

As a final remark, more accurate measurements of the QLF faint-end and its redshift evolution can be addressed with the much wider low-luminosity quasar survey with the HSC-SSP.
\acknowledgments
MO would like to express gratitude to Linhua Jiang for kindly providing his $z\sim6$ QLF data based on the SDSS.


The Hyper Suprime-Cam (HSC) collaboration includes the astronomical
communities of Japan and Taiwan, and Princeton University.  The HSC
instrumentation and software were developed by the National
Astronomical Observatory of Japan (NAOJ), the Kavli Institute for the
Physics and Mathematics of the Universe (Kavli IPMU), the University
of Tokyo, the High Energy Accelerator Research Organization (KEK), the
Academia Sinica Institute for Astronomy and Astrophysics in Taiwan
(ASIAA), and Princeton University.  Funding was contributed by the FIRST 
program from Japanese Cabinet Office, the Ministry of Education, Culture, 
Sports, Science and Technology (MEXT), the Japan Society for the 
Promotion of Science (JSPS),  Japan Science and Technology Agency 
(JST),  the Toray Science  Foundation, NAOJ, Kavli IPMU, KEK, ASIAA,  
and Princeton University.


This work was supported by JSPS KAKENHI Grant Numbers JP15J02115, JP15H03645, JP17H04830.
M.I. acknowledges the support from the NRFK grant No.~2017R1A3A3001362.

%

\vspace{5mm}
\facilities{Subaru}

\end{document}